\documentclass{article}

\usepackage{arxiv}

\usepackage[utf8]{inputenc} 
\usepackage[T1]{fontenc}    
\usepackage{hyperref}       
\usepackage{url}            
\usepackage{booktabs}       
\usepackage{amsfonts}       
\usepackage{nicefrac}       
\usepackage{microtype}      
\usepackage{lipsum}
\usepackage{soul,color}
\usepackage{graphicx}
\usepackage{multirow}
\usepackage{caption}
\usepackage{commath}
\usepackage{multicol}
\usepackage[numbib]{tocbibind}
\usepackage{algorithm}
\usepackage{float}
\usepackage{caption}
\usepackage{subcaption}
\usepackage[table,xcdraw]{xcolor}

\usepackage{wrapfig}
\usepackage{fancyhdr}       
\usepackage{graphicx}       
\usepackage{float}
\usepackage{amsmath}

\usepackage[section]{placeins}
\usepackage{algpseudocode}
\graphicspath{{figures/}}     

\title{Genetic programming-based learning of carbon interatomic potential for materials discovery}

\author{
  Andrew Eldridge \\
  Department of Computer Science and Engineering\\
  University of South Carolina\\
  Columbia, SC, 29201, USA \\
\And
 Alejandro Rodriguez \\
  Department of Mechanical Engineering\\
  University of South Carolina\\
  Columbia, SC, 29208, USA \\
   \And
 Ming Hu* \\
  Department of Mechanical Engineering\\
  University of South Carolina\\
  Columbia, SC, 29208, USA \\
  \texttt{hu@sc.edu} \\
   \And
 Jianjun Hu \thanks{*Corresponding authors. Tel: 803-777-7304. E-mail: Jianjunh@cse.sc.edu (Jianjun Hu) and Hu@sc.edu (Ming Hu)}\\
  Department of Computer Science and Engineering\\
  University of South Carolina\\
  Columbia, SC, 29201, USA \\
  \texttt{jianjunh@cse.sc.edu} \\
}

\begin{document}
\maketitle

\begin{abstract}
Efficient and accurate interatomic potential functions are critical to computational study of materials while searching for structures with desired properties. Traditionally, potential functions or energy landscapes are designed by experts based on theoretical or heuristic knowledge. Here, we propose a new approach to leverage strongly typed parallel genetic programming (GP) for potential function discovery. We use a multi-objective evolutionary algorithm with NSGA-III selection to optimize individual age, fitness, and complexity through symbolic regression. With a DFT dataset of 863 unique carbon allotrope configurations drawn from 858 carbon structures, the generated potentials are able to predict total energies within  $\pm 7.70$ eV at low computational cost while generalizing well across multiple carbon structures. Our code is open source and available at \url{http://www.github.com/usccolumbia/mlpotential}.
\end{abstract}

\keywords{Genetic programming \and symbolic regression \and Pareto optimization \and machine learning potential}

\section{Introduction}
\label{sec:introduction}

Machine learning (ML) potentials have been prevalent in computational materials science in molecular dynamics (MD) simulations for several decades \cite{li2020unified,rowe2020accurate,deringer2020general,behler2016perspective} and have increasingly demonstrated their ability to obtain results approaching ab initio accuracy without the associated complexity of calculations from first principles. However, training accurate and transferable machine learning potentials remains to be a challenging problem \cite{rowe2020accurate}. Currently, there are three main categories of machine learning potentials for materials simulation including neural networks, kernel models, and genetic programming derived analytical expressions.

As one of the most promising machine learning potentials, neural network potentials (NNPs), initially only effective in small systems, have recently been able to simulate systems containing thousands of atoms with near density-functional theory (DFT) levels of accuracy \cite{HighDimNNPotentials}. Whereas early neural network potentials considered a limited subset of short-range local interactions, the latest high-dimensional neural network potentials (HDNNPs) have additionally accounted for non-local interactions and charge transfer, resulting in improved accuracy for similar computational cost \cite{FourthGenHighDimNNPotentions}. Recently, HDNNPs have even been able to perform MD simulations of a 100 million-atom copper system over nanosecond intervals with ab initio accuracy on the Summit supercomputer \cite{DeepPotentialMolecularDynamicsSimulation100mil}. Several physical properties of atomic systems have been approximated to a high degree of accuracy by NNPs, including lattice thermal conductivity \cite{li2020unified}, solvation energy \cite{ChemicalPropertiesPredictionNNP}, density, porosity, and hardness \cite{AluminumCopperCompositeMatPotential}. Neural network potentials generally perform better than ML models based on regression of parameters in a heuristically designed potential function with respect to both accuracy and transferability \cite{PhysicallyInformedNNP} at the cost of a black-box model. The inherent black-box design of NNPs is one of their greatest limitations, making analysis of the physical properties considered by these models very difficult. This issue is addressed by regression models, which provide a white-box model by producing intelligible mathematical expressions consisting of scaling parameters and kernel functions to describe the physical properties of a system.

Kernel-based regression, where parameters in a heuristically designed potential with fixed kernels are programmatically optimized, is another popular method for ML potential development \cite{ramakrishnan_von_lilienfeld_2015, Scherer2020, MLPotentialsMatterSim, EmpiricalApproachEnergyCovalentSystems, MLInteratomicPotentialAmorphousCarbon}. Potentials developed through kernel-based regression have achieved near-DFT accuracy with heuristically designed terms for two-, three-, and many-body interactions \cite{EmpiricalApproachEnergyCovalentSystems}. For instance, Gaussian approximation potential (GAP) implementations of kernel-based regression have been able to calculate energy within hundredths of eV/atom relative to DFT target calculations \cite{MLInteratomicPotentialAmorphousCarbon}. The GAP produced by this research expresses total energy as a sum of two-, three-, and many-body contributions, shown in equation~\eqref{eq:gap_total_energy}.

\begin{equation}
\label{eq:gap_total_energy}
\begin{split}
    E = \left( \delta^{(2b)} \right)^2 \sum_{i \in \text{pairs}} \epsilon^{(2b)} \left( q_i^{(2b)} \right) + \\
    \left( \delta^{(3b)} \right)^2 \sum_{j \in \text{triplets}} \epsilon^{(3b)} \left( q_j^{(3b)} \right) + \\
    \left( \delta^{(MB)} \right)^2 \sum_{a \in \text{atoms}} \epsilon^{(MB)} \left( q_a^{(MB)} \right)
\end{split}
\end{equation}

Each descriptor is assigned a local energy contribution according to the kernel defined in equation~\eqref{eq:gap_local_energy_kernel}.

\begin{equation}
\label{eq:gap_local_energy_kernel}
    \epsilon^{(d)}\left( q^{(d)} \right) = \sum_{t=1}^{N_t^{(d)}} \alpha_t^{(d)} \kappa^{(d)} \left( q^{(d)}, q_t^{(d)} \right)
\end{equation}

where $q^{(d)}$ is a vector representation of a given descriptor.

A squared exponential kernel, shown in equation~\eqref{eq:gap_squared_exponential_kernel}, is used to represent the contributions of two- and three- body interactions, and a Smooth Overlap of Atomic Potentials (SOAP) \cite{SOAP} kernel is used for many-body interactions.

\begin{equation}
\label{eq:gap_squared_exponential_kernel}
    \kappa^{(d)}\left( q_i^{(d)}, q_t^{(d)} \right) = \exp{\left[ - \frac{1}{2} \sum_\xi \frac{\left( q_{\xi,i}^{(d)} - q_{\xi,t}^{(d)} \right)^2}{\theta_\xi^2} \right]}
\end{equation}

These are only a few examples drawn from the complete model. There are many more nested kernels, making for a highly complex model of a strictly defined form. For a complete description of the GAP, refer to ref.~\cite{MLInteratomicPotentialAmorphousCarbon}.

The use of these kernels enables the model to obtain high accuracy by modifying a small number of parameters, but presents two significant limitations in the model: high complexity and inflexibility. The nesting of multiple kernels, many of which contain summations over neighbors or triplets of atoms, in kernel-based regression models makes for potentially unnecessary computational expense as demonstrated by spectral neighbor analysis potentials (SNAP), which have reduced the computational expense of force calculations by an order of magnitude relative to GAPs by assuming a previously unrecognized linear relationship between atomic energy and bispectrum components \cite{SpectralNeighborPotential}. Further unrecognized relationships could be discovered by GP-based regression potentials through the process of symbolic regression, whereas traditional kernel-based models are unable to improve upon their basic form without further heuristic contributions from expert analysis.

Here we propose a method of genetic programming-based symbolic regression (GPSR) to develop interatomic potentials in analytical forms with high accuracy and transferability. This approach differs from kernel-based regression methods in its capacity to "rediscover" fundamental properties of atomic systems without the restriction of a strictly-defined kernel imposed on the model. Theoretically, this allows GPSR algorithms to outperform their kernel-based equivalents given substantial time and training data. The GPSR framework used here can additionally be applied to non-carbon DFT datasets to generate a variety of interatomic potentials for any type of many-body system, making GPSR models highly transferable. This transferability of the GPSR framework is one of its most notable attributes, removing the need to modify the model's basic form in order to develop potentials for multiple diverse systems. GPSR potential functions also have the advantage of high interpretability with their analytical forms of functions \cite{lensen2020genetic}.

\begin{figure}[ht!]
  \centering
  \includegraphics[width=0.75\textwidth]{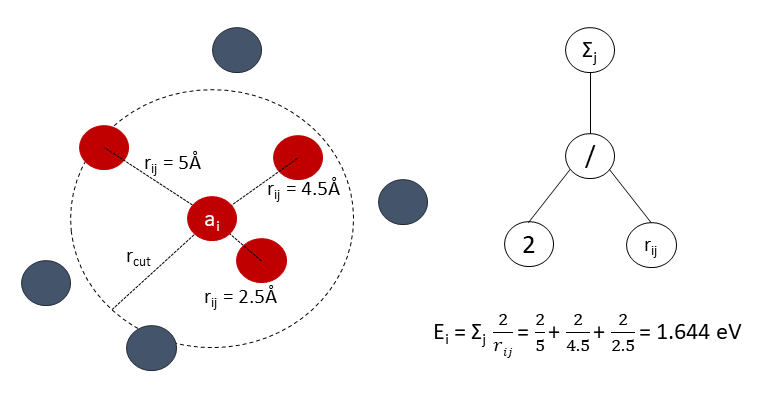}
  \caption{Example of neighboring atom pairs considered in two-body interactions (left) and an example expression tree (right). The red atoms include the central atom $a_i$ and its neighbors with a Euclidean distance less than $r_{cut}$ from the central atom. The blue atoms are non-neighboring. The GPSR framework presented here considers strictly two-body interactions to develop rudimentary proof-of-concept potentials. The local two-body energy contribution $E_i$ is derived as above from the potential's expression tree and atom configuration.}
  \label{fig:atomic_neighbors}
\end{figure}

The GPSR framework for potential function development is not a novel concept. Research into this method of potential development over the past few years has demonstrated the ability of GPSR models to replicate known functional forms with high accuracy \cite{wang_wagner_rondinelli_2019, PotentialsSymbolicRegression, sun_ouyang_zhang_zhang_2019}. One such model was used to develop low-complexity potentials using a relatively homogeneous DFT dataset of 32-atom copper structures with a minimum MAE of 3.68 meV/atom in their energy calculations \cite{PotentialsSymbolicRegression}. This research was limited to a dataset of 150 energies and didn't demonstrate the ability of produced regression models to generalize well to structures dissimilar to the training set. Here we use a carbon dataset of 863 energies calculated by DFT and demonstrate the ability of generated models to generalize with multiple train-test partitions. We have also modified elements of the evolutionary algorithm in order to minimize premature convergence through multiobjective age-fitness Pareto optimization.

Here, we consider only two-body interactions to serve as a proof of concept. Two-body interactions are defined between any given atom $a_i$ in the system and all neighboring atoms $a_j$ contained in the sphere centered about $a_i$ with radius $r_{cut}$, visualized in Figure~\ref{fig:atomic_neighbors}.

One drawback of symbolic regression is the exceptionally large hypothesis space. Thus any regression algorithm operating in this space should be able to quickly differentiate between viable and unviable solutions while avoiding premature convergence to a suboptimal solution. Genetic programming-based implementations address this concern with their exceptional ability to quickly explore large hypothesis spaces and identify high-fitness individuals. Furthermore, strongly-typed GP greatly limits the hypothesis space by imposing restrictions on what combinations of operators and operands constitute a valid individual \cite{10.1162/evco.1995.3.2.199}.

The potentials produced here achieved a minimum RMSE of 6.915 eV for their energy calculations. They do not yet possess the accuracy of kernel-based GAPs, which have achieved RMSE scores as low as 0.002 eV/atom for certain 125-atom crystalline carbon structures \cite{MLInteratomicPotentialAmorphousCarbon}. This is expected. The non-trivial relationship between atomic structure and formation energy necessitates the use of heuristic knowledge based on first principles for regression methods to approximate DFT accuracy with a limited training set. Furthermore, the GPSR framework produced here only considers two-body interactions. We predict that the introduction of three- and many-body terms to the framework will greatly improve the accuracy of generated potentials, which has been demonstrated in related GAP research \cite{MLInteratomicPotentialAmorphousCarbon}. With improvements in data acquisition methods and the increased availability of large amounts of ab initio data \cite{MLPotentialsMatterSim}, strongly typed GP-based regression models have the potential overtake heuristically designed models in both accuracy and computational complexity in a manner analogous to HDNNPs without the associated downside of a black-box model.

\section{Genetic programming overview}
\label{sec:gpoverview}

Genetic programming (GP) simulates the biological process of natural selection in a computer program. An evolutionary algorithm is applied to individuals, or members of a population, across multiple generations. A generation is defined as a group of individuals being evaluated simultaneously by the algorithm; the offspring of a generation are the individuals produced by one or more individuals in that generation.

\subsection{Genetic operations}
\begin{figure}%
    \centering
    \begin{subfigure}[t]{0.49\textwidth}
        \includegraphics[width=7.5cm]{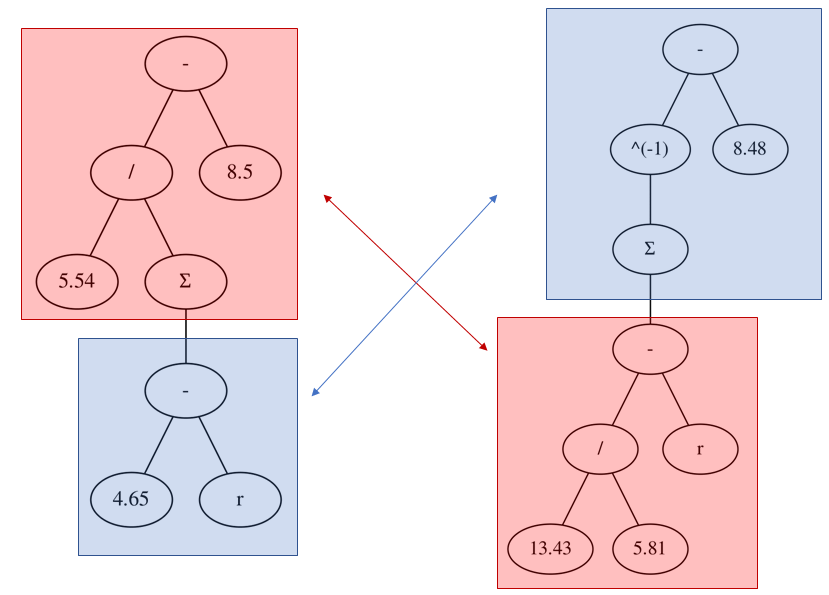}
        \caption{}
        \vspace{-3pt}
        \label{fig:crossover}
    \end{subfigure}
    \begin{subfigure}[t]{0.49\textwidth}
        \includegraphics[width=7.5cm]{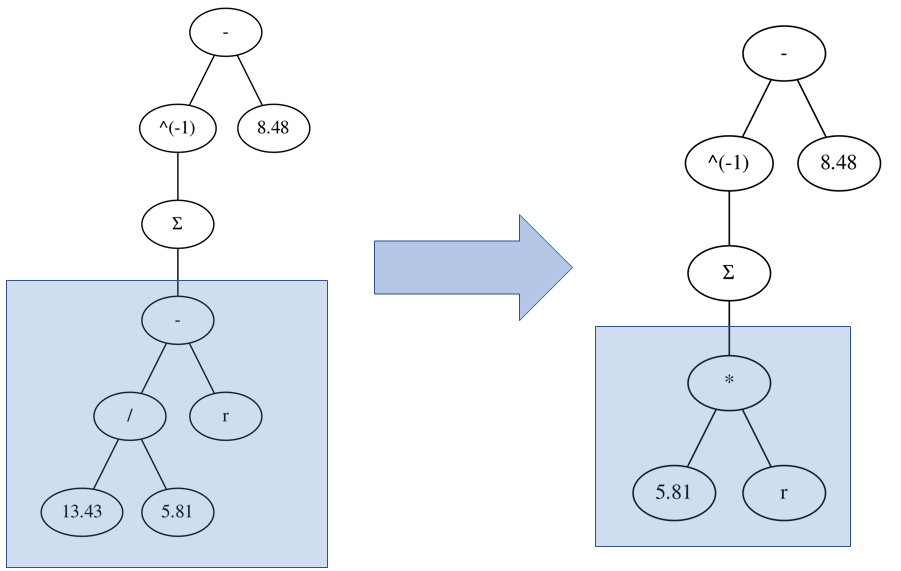}
        \caption{}
        \vspace{-3pt}
        \label{fig:mutation}
    \end{subfigure} 
    \caption{Genetic operations for crossover (left) and mutation (right). (a) Example of a crossover operation. Subsets of each parents' genes are sampled to produced two offspring (one blue, one red). (b) Example of a mutation operation. The parent (left) has a subset of its genes mutated to produce an offspring (right).}%
    \label{fig:genetic_operations}%
\end{figure}

The evolutionary process consists of two genetic operations: crossover and mutation. The crossover operation combines random subsections of two individuals to produce a new offspring which is a combination of those two individuals. The mutation operation replaces a subset of a single individual with some randomly generated expression tree. Examples of both crossover and mutation operations are shown in Figure~\ref{fig:genetic_operations}.

\subsection{Evaluation function}
The evaluation (or objective) function evaluates the fitness of an individual as an $n$-tuple, where $n$ is the number of fitness objectives defined. The evaluation function is applied to the initial population and all successive generations when determining individual fitness. In our case, there are three fitness objectives: energy calculation RMSE, individual age, and individual complexity. These are addressed in detail in section~\ref{sec:implementation}.

\subsection{Selection algorithm}
\begin{figure}
  \centering
  \includegraphics[width=0.75\textwidth]{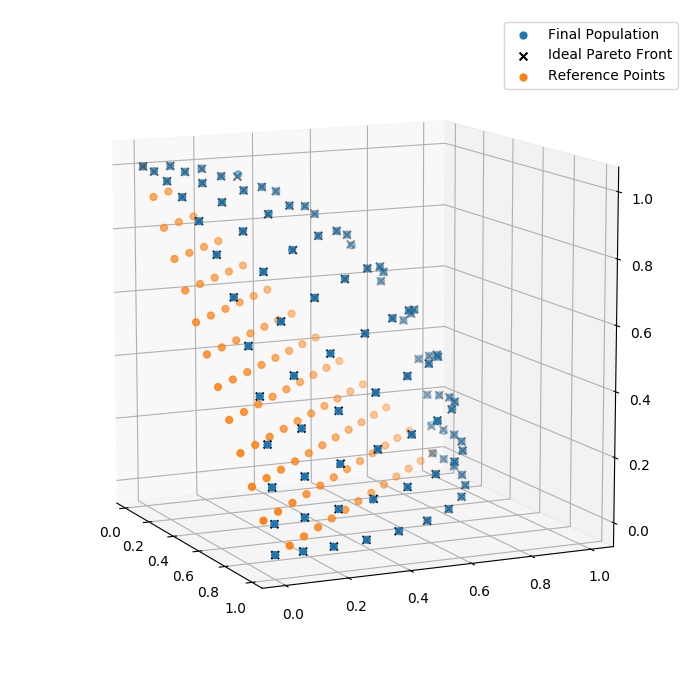}
  \caption{Example of Pareto front generated using NSGA-III selection in DEAP \cite{deap_nsgaiii_image}. Individuals along the front are non-dominated and estimate the target values (ideal Pareto front) with high accuracy. Reference points are fixed points defined on initialization of the NSGA-III algorithm \cite{deap}.}
  \label{fig:nsga3_pareto_front}
\end{figure}

The selection algorithm determines which individuals from the union of the current population and its offspring to preserve into the following generation. Here, we use the Non-dominated Sorting Genetic Algorithm III (NSGA-III) for multiobjective selection. Figure~\ref{fig:nsga3_pareto_front} shows a sample set of reference points, target points, and Pareto front of non-dominated individuals produced by an NSGA-III implementation in DEAP \cite{deap}.

\subsection{Evolutionary algorithm}
The evolutionary algorithm contains the main logic loop of the evolutionary process. Individuals in an initial population are evaluated by the evaluation function. The cumulative best individual after each generation is preserved in the "hall of fame." Offspring are produced by applying the crossover and mutation operations to the current population. Here we use a modified version of DEAP's EA simple algorithm \cite{deap} with elements of age-fitness Pareto optimization \cite{AgeFitness}, where each individual's age is derived from it's parents' ages depending on the genetic operation used to produce the individual; each individual's age is then incremented for each generation it remains in the population. Finally, the selection algorithm is used to select which individuals will be preserved into the following generation. Pseudocode for the evolutionary algorithm is provided in Algorithm~\ref{alg:age_fitness}.

\begin{algorithm}
\caption{A simple age-fitness evolutionary algorithm. Required parameters include a population size ($\mu$) and initial population (pop). In each generation, offspring are produced and evaluated, then a selection algorithm is applied to the whole population to retrieve an updated population. Each inidividual's age is incremented for every generation it survives.}
\label{alg:age_fitness}
\begin{algorithmic}[1]
\Require $\mu \geq 0$, pop
\State evaluated $\gets$ eval(pop)
\State hof $\gets$ lowestErrorInd(evaluated)

\State gen $\gets$ 0
\While{gen $< \mu$}
\State offspring $\gets$ applyGeneticOps(pop)
\State evaluated $\gets$ eval(offspring)
\State hof $\gets$ lowestErrorInd(evaluated + hof)

\State $i \gets 0$
\While{$i <$ len(pop)}
\State pop[i].age $\gets$ pop[i].age + 1
\State $i \gets i+1$
\EndWhile

\State pop $\gets$ select(pop + offspring, $\mu$)

\State gen $\gets$ gen+1
\EndWhile
\end{algorithmic}
\end{algorithm}

\subsection{Hyperparameters}
The GP model is characterized by several hyperparameters which are crucial to its performance. The mutation probability ($P_{mut}$) and crossover probability ($P_{cx}$) define the probability of a crossover operation between two individuals and a mutation of a single individual, respectively.

The population size ($\mu$) and the number of generations to evaluate ($N_{gen}$) contribute to the runtime of the evolutionary algorithm and affect how the hypothesis space is explored. For instance, an instance characterized by relatively high $\mu$ and low $N_{gen}$ would explore a larger breadth of the hypothesis space with each generation. Conversely, an instance characterized by low $\mu$ and high $N_{gen}$ would explore a greater depth of solutions, as more genetic crossovers and mutations would occur with a greater number of generations. These two parameters must be balanced to explore the hypothesis space effectively.

The minimum and maximum depth of new individuals generated are specified by $D_{min}$ and $D_{max}$, respectively. The mutation operation is parameterized by the minimum and maximum depth of the subtree to be mutated ($SD_{min}$ and $SD_{max}$, respectively), as well as the minimum and maximum depth of mutation trees to replace the mutated region ($MD_{min}$ and $MD_{max}$, respectively).

The hyperparameters used in our implementation are shown in Table ~\ref{tab:hyperparams} 

It's worth noting that these parameters were not systematically optimized. Rather, our hyper-parameters were selected heuristically based on a limited number of validation runs. Systematic optimization of these parameters, which is a computationally expensive process, would likely produce more accurate regression models and reduced runtime. For further discussion on this, see future work in section~\ref{sec:future_work}.

\begin{table}
    \centering
    \caption{The GP hyperparameters used to train the GP7 model.}
    \begin{tabular}{|l|l|}
     \hline
     \textbf{Name} & \textbf{Value}\\
     \hline
     $P_{mut}$ & 0.1\\
     $P_{cx}$ & 0.7\\
     $\mu$ & 2500\\
     $N_{gen}$ & 1000\\
     $D_{min}$ & 1\\
     $D_{max}$ & 13\\
     $SD_{min}$ & 1\\
     $SD_{max}$ & 8\\
     $MD_{min}$ & 1\\
     $MD_{max}$ & 8\\
     \hline
    \end{tabular}
    \label{tab:hyperparams}
\end{table}

\section{Implementation details}
\label{sec:implementation}

Originally, the FastSR \cite{FastSR} library for symbolic regression and FastGP \cite{FastGP} library for genetic algorithms were used to develop the model. However, the summation operation in this particular symbolic regression implementation is non-trivial to develop in FastSR's highly abstracted API. Furthermore, FastGP only supports loosely typed genetic programming. In the end, we opted to develop a custom evolutionary algorithm using the DEAP API \cite{deap}.

We used a multiobjective evolutionary algorithm with NSGA-III selection and defined fitness as a 3-tuple ($\alpha, \beta, \gamma$), where $\alpha$ is the root mean squared error of an individual's energy calculation, $\beta$ is an individual's age, and $\gamma$ is an individual's complexity, measured in number of nodes.

NSGA-III selection includes fixed-rate crossover and mutation operations; our training metaparameters include $P_{CX} = 0.7$ for crossover probability and $P_{MUT} = 0.1$ for mutation probability. NSGA-III also includes elitism, where nondominated individuals from each generation are preserved into the following generation independent of the usual selection process, thereby avoiding devolutions while maintaining a diverse population \cite{nsgaIII}.

Our evolutionary algorithm implements components of an age-fitness Pareto optimization algorithm used to prevent premature convergence to a local optimum by assigning individuals an age value based on the max of their parents' age and the operation used to generate the individual (crossover or mutation) \cite{AgeFitness}. DFT calculations are used as the base truth for training, and all fitness parameters are assigned negative weights as this is a minimization problem on all fronts.

\subsection{Hypothesis space}
Individuals in the hypothesis space are defined:

\[
f : \mathbb{R}_{\geq 0}^n \to \mathbb{R}
\]
\[
r_i \mapsto E_i
\]

Where $r_i$ is a vector of $n$ interatomic distances between the $i$-th atom and its neighbors, and $E_i$ is the net energy attributed to the $i$-th atom's two-body interactions. A neighbor of the $i$-th atom is defined to have a Euclidean distance less than 5{\AA} from the $i$-th atom (i.e. $r_{cut} = 5${\AA}). The total energy, representing our GP objective function, is then derived as a summation of local energy contributions, shown in equation \eqref{eq:GP_objective}.

\begin{equation} \label{eq:GP_objective}
    E = \sum_{i=1}^n E_i = \sum_{i=1}^n f \circ r_i
\end{equation}

\subsection{GP Primitive set}
We utilized strongly typed genetic programming to run the genetic algorithm on a strictly defined set of primitives and terminal values. A strongly typed GP implementation was necessary due to type-restricted operations (e.g. a square root node's input must be a positive float) and the presence of mixed vector and scalar operations in the primitive set. Without strongly typed GP, the hypothesis space would be flooded with syntactically invalid individuals.

The GP primitive set consists of basic arithmetic operations, inversion, summation, simple powers and roots, an ephemeral constant, and a single argument ($r_i$) which describes the interatomic distance ({\AA}) between neighboring atoms (see Table ~\ref{tab:primitiveset}).

Scalar and vector variants of each operation exist in the primitive set. In general, scalar operations follow the form $\mathbb{R} \to \mathbb{R}$ or $\mathbb{R} \to \mathbb{R} \to \mathbb{R}$ for 1- and 2-arity operations, respectively. Vector operations generally follow the form $\mathbb{R}^n \to \mathbb{R}^n$ or $\mathbb{R}^n \to \mathbb{R}^n \to \mathbb{R}^n$ for 1- and 2-arity operations, respectively. The summation operation uniquely maps a vector to a scalar: $\mathbb{R}^n \to \mathbb{R}$.

\subsection{Parallelization}
Our evolutionary algorithm contains multiprocessing support for parallel evaluation of individuals in the same generation. We used the Hyperion high performance computing (HPC) cluster at the University of South Carolina to train the model across 40 2.8GHz nodes simultaneously. The best performing model, GP7, was trained for a total of 45.5 hours on this cluster.

\begin{table}[ht]
    \centering
    \caption{The complete GP Primitive set. Every scalar operation listed has a vector equivalent. The final output of a GP tree is in $\mathbb{R}$, corresponding to the predicted local energy contribution $E_i$. Thus, terminals and operations from the primitive set must be combined in such a way that an input vector $r_i \in \mathbb{R}_{\geq 0}^n$ produces a scalar $f(r_i) \in \mathbb{R}$.}
    \def\arraystretch{1.5}
    \begin{tabular}{|l|l|l|}
     \hline
     \textbf{0-arity (terminal)} & \textbf{1-arity} & \textbf{2-arity}\\
     \hline
     Ephemeral constant [0.0, 30.0) : $\mathbb{R}_{\geq 0}$ & Square : $\mathbb{R} \to \mathbb{R}_{\geq 0}$ & Addition : $\mathbb{R} \to \mathbb{R} \to \mathbb{R}$\\
     Interatomic distances ($r_i$) : $\mathbb{R}_{\geq 0}^n$ & Square root : $\mathbb{R}_{\geq 0} \to \mathbb{R}_{\geq 0}$ & Subtraction : $\mathbb{R} \to \mathbb{R} \to \mathbb{R}$\\
     & Cube root : $\mathbb{R} \to \mathbb{R}$ & Multiplication : $\mathbb{R} \to \mathbb{R} \to \mathbb{R}$\\
     & Inversion : $\mathbb{R} \to \mathbb{R}$ & Division : $\mathbb{R} \to \mathbb{R} \to \mathbb{R}$\\
     & Absolute value : $\mathbb{R} \to \mathbb{R}_{\geq 0}$ & \\
     & Summation : $\mathbb{R}^n \to \mathbb{R}$ & \\
     \hline
    \end{tabular}
    \label{tab:primitiveset}
\end{table}

\section{Results}
\label{sec:results}

The genetic algorithm presented here demonstrates the ability of GP-based regression models to rapidly identify DFT approximations across a diverse set of carbon configurations with lower computational complexity than existing GAPs. The trained models generalize well across multiple dataset partitions, indicating the successful rediscovery of underlying physical properties.

\subsection{Model accuracy and transferability}

\begin{figure}[ht!]%
    \centering
    \begin{subfigure}[t]{0.49\textwidth}
        \includegraphics[width=7.5cm]{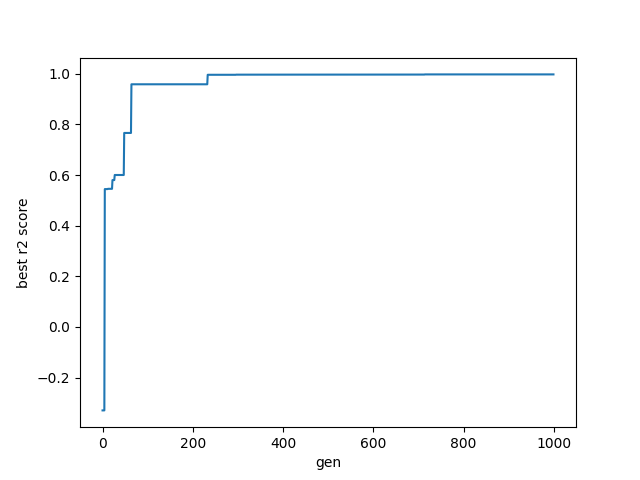}
        \caption{$r^2$ score progression}
        \vspace{-3pt}
        \label{fig:r2}
    \end{subfigure} 
    \begin{subfigure}[t]{0.49\textwidth}
        \includegraphics[width=7.5cm]{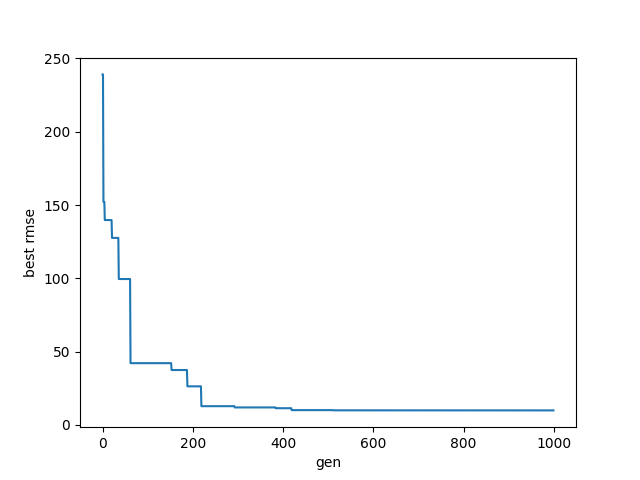}
        \caption{RMSE score progression}
        \vspace{-3pt}
        \label{fig:rmse}
    \end{subfigure} 
    \caption{The 1000-generation progression of $r^2$ and RMSE scores for GP2. Incremental improvements in both metrics are shown to occur simultaneously throughout the evolutionary process, with early generations making the largest and most abrupt improvements in the model. Later generations involve smaller and less frequent modifications to the best individual as the space of improvements over the current model becomes smaller and harder to identify.}%
    \label{fig:gp2_gen_metrics}%
\end{figure}

The evolutionary process of our potentials was characterized by a rapid convergence to pseudo-optimal solutions within the first few hundred generations, followed by slower incremental improvements in the later generations. Figure~\ref{fig:gp2_gen_metrics} shows the progression of the $r^2$ and RMSE scores of the highest performing individual over time in the GP instance which produced GP2. After only 1000 generations, an $r^2$ of 0.997 and RMSE of 9.951 were obtained with a complexity of 14 nodes, evaluated in linear time with no nested summations. The pattern of incremental improvements exhibited in the figure is typical of genetic processes, as improvements are not made with every generation and they can be of varying significance. Typically, the largest improvements are seen in early generations and the fitness curve begins to flatten as the number of generations grows.

\begin{figure}[ht!] 
    \begin{subfigure}[t]{0.49\textwidth}
        \includegraphics[width=\textwidth]{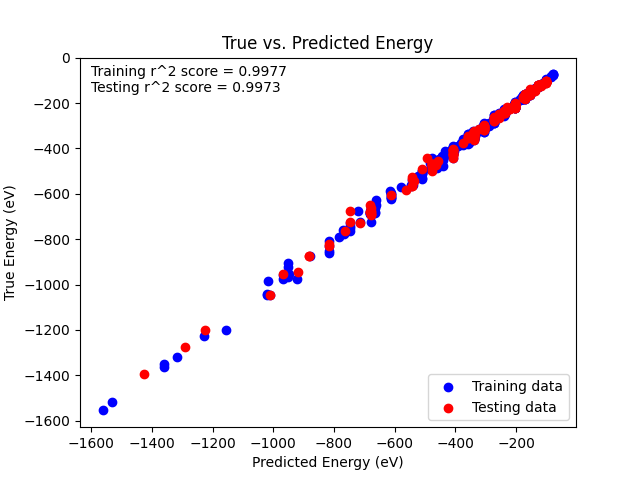}
        \caption{}
        \vspace{-3pt}
        \label{fig:gp3_predict}
    \end{subfigure}
    \begin{subfigure}[t]{0.49\textwidth}
        \includegraphics[width=\textwidth]{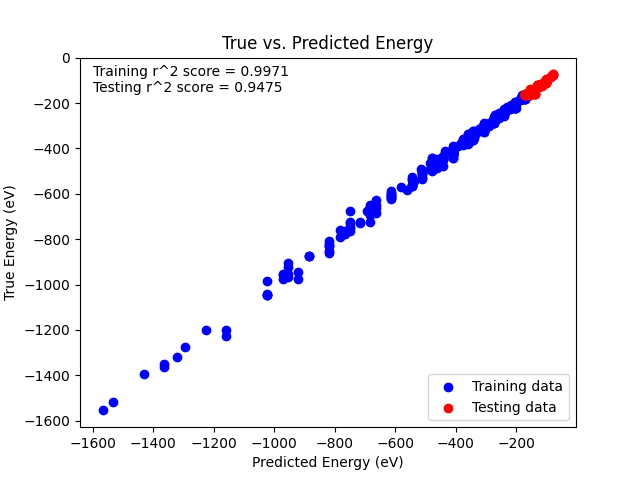}
        \caption{}
        \vspace{-3pt}
        \label{fig:gp5_predict}
    \end{subfigure} 
 \begin{subfigure}[t]{0.50\textwidth}
        \includegraphics[width=\textwidth]{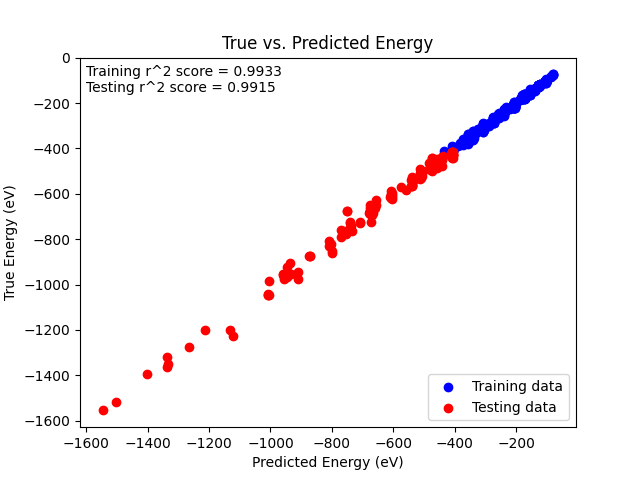}
        \caption{}
        \vspace{-3pt}
        \label{fig:gp6_predict}
    \end{subfigure}              
    \begin{subfigure}[t]{0.50\textwidth}
        \includegraphics[width=\textwidth]{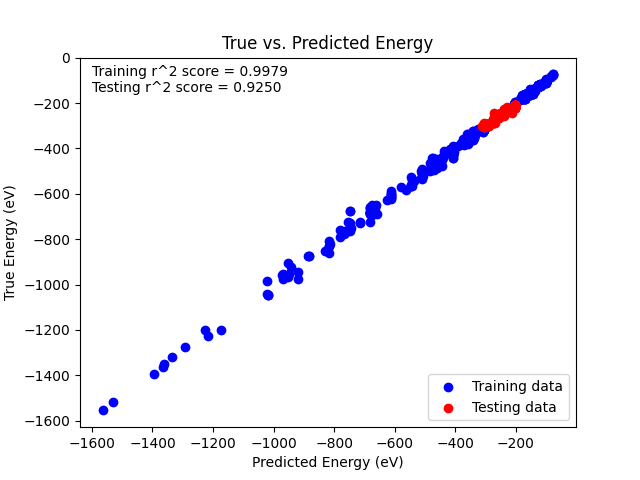}
        \caption{}
        \vspace{-3pt}
        \label{fig:gp7_predict}
    \end{subfigure}     

   \caption{Predictive performance of generated models relative to target DFT energies. Across all four partitions, the GP-based models performed with similar accuracy and high correlation to target energies ($r^2 > 0.99$). From the models' performances in each scenario, we conclude that the produced regression models generalize well to novel carbon structures, suggesting the identification of underlying physical properties by the models. (a) GP3 energy predictions (random train-test partition). (b) GP5 energy predictions (80/20 train-test partition). (c) GP6 energy predictions (20/80 train-test partition). (d) GP7 energy predictions (40/20/40 train-test partition). }
  \label{fig:predictionplots}
\end{figure}

All selected potentials generalized well across a variety of train-test partitions of the heterogeneous dataset of 863 carbon configurations drawn from 858 unique structures. The $r^2$ scores of top-performing models exceeded 0.99 on both training and testing sets, shown in Figure~\ref{fig:predictionplots}. This is particularly notable in the non-random partitions (80-20, 20-80, 40-20-40), which demonstrate the ability of the generated regression models to generalize well to carbon structures dissimilar to those included in the training data. This indicates that the evolutionary process is rediscovering fundamental underlying properties rather than fitting specifically to the testing dataset.

\subsection{Generated potentials}

\begin{figure}[ht!] 
    \begin{subfigure}[t]{0.49\textwidth}
        \centering
        \includegraphics[width=0.6\textwidth]{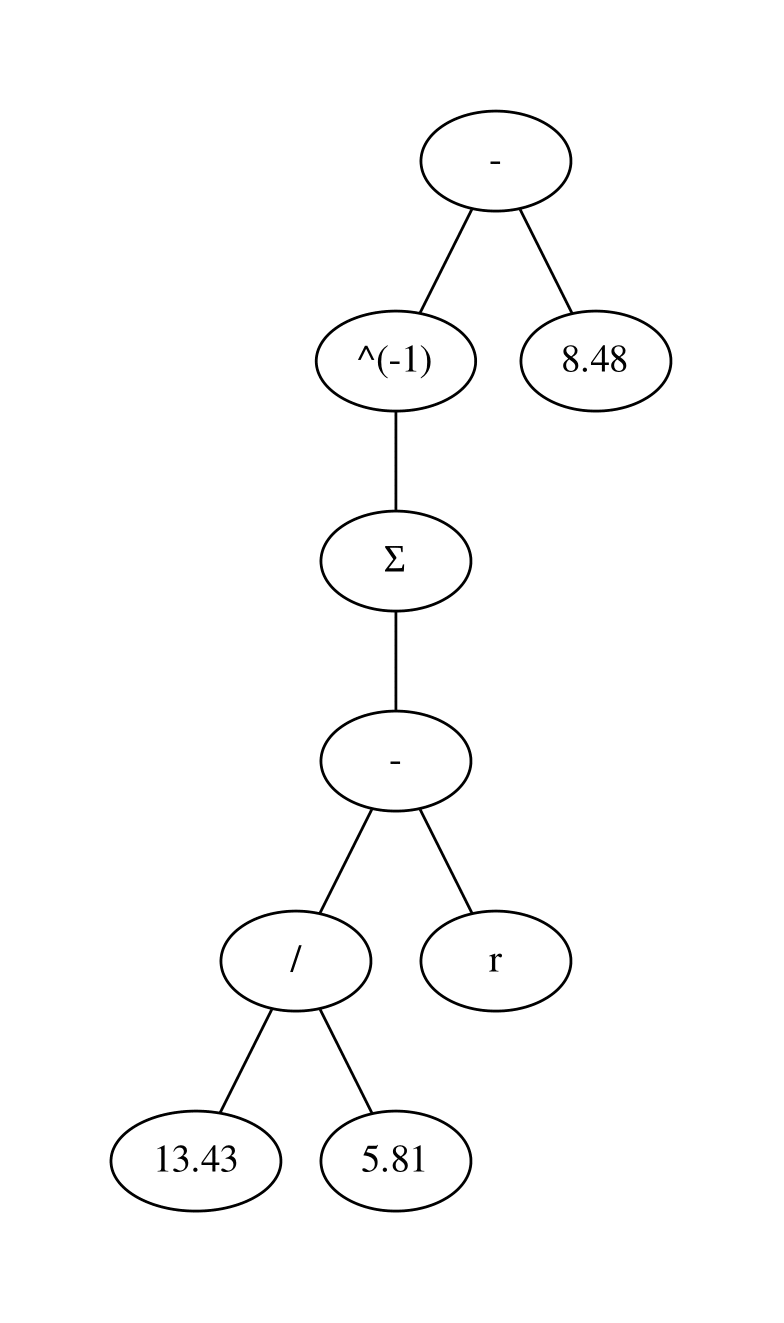}
        \caption{GP1}
        \vspace{-3pt}
        \label{fig:mixChart}
    \end{subfigure}
    \begin{subfigure}[t]{0.49\textwidth}
        \centering
        \includegraphics[width=0.6\textwidth]{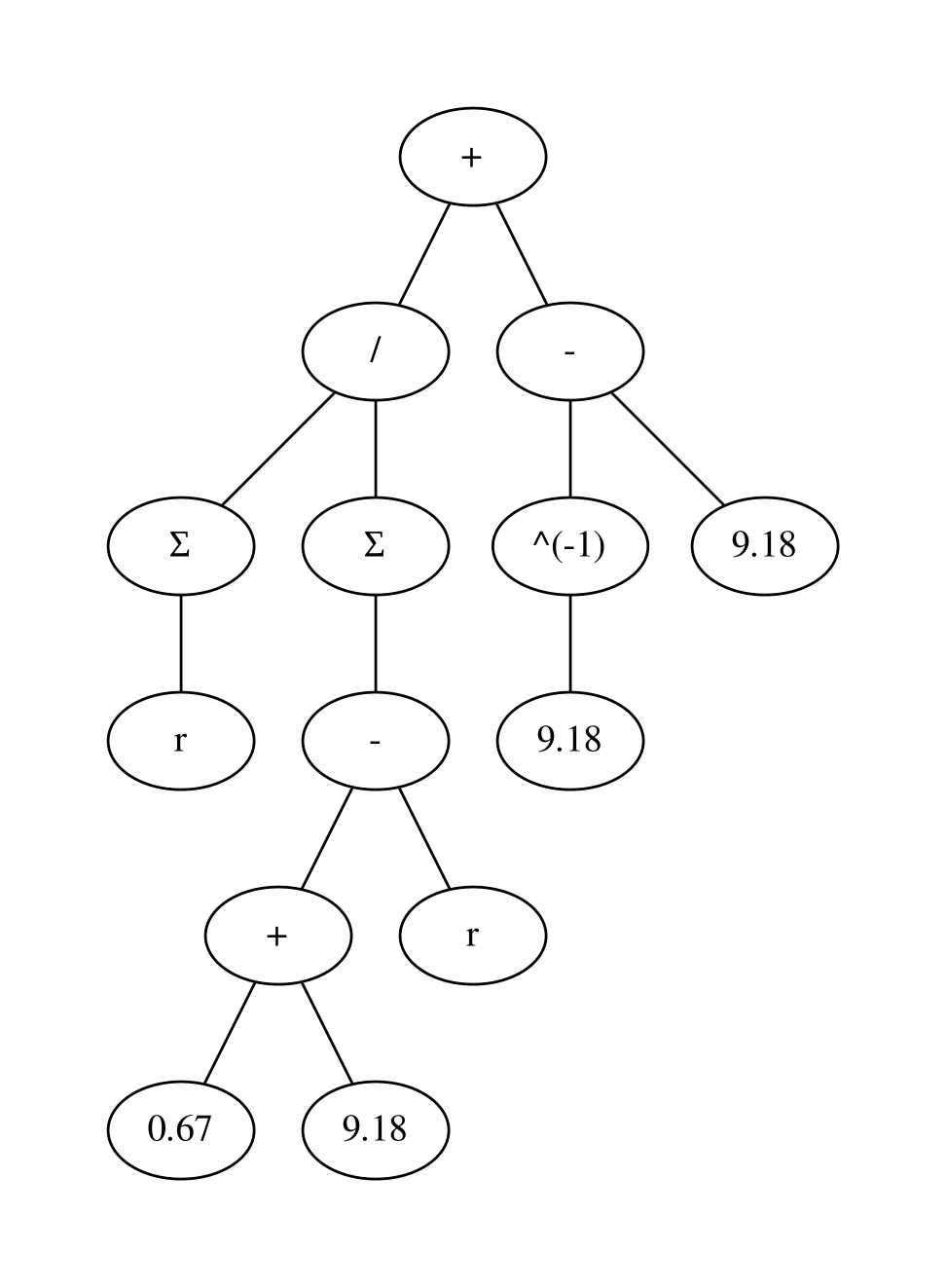}
        \caption{GP2}
        \vspace{-3pt}
        \label{fig:pureChart}
    \end{subfigure} 
 \begin{subfigure}[t]{0.50\textwidth}
        \centering
        \includegraphics[width=0.6\textwidth]{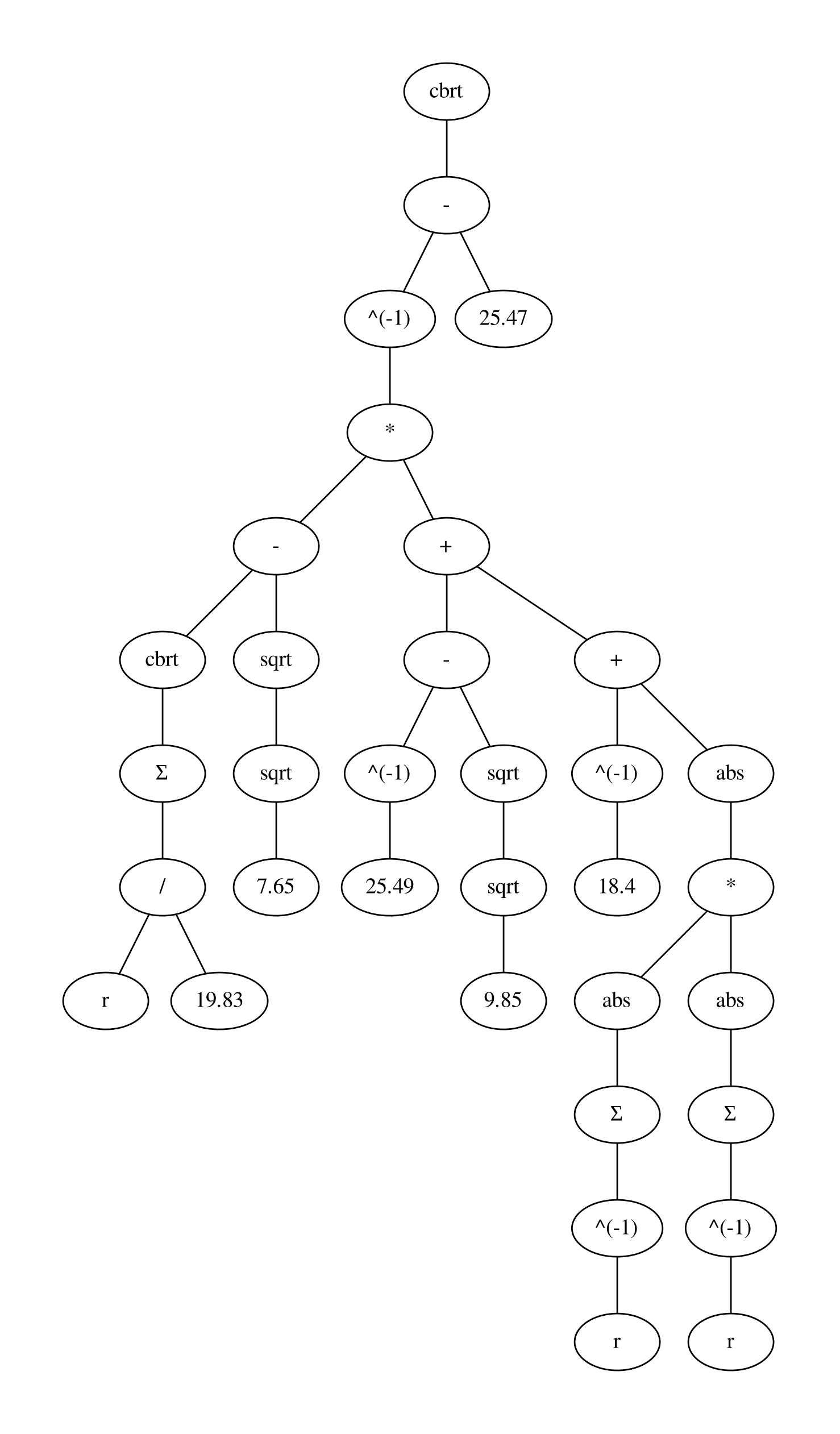}
        \caption{GP3}
        \vspace{-3pt}
        \label{fig:GaB2N3_target}
    \end{subfigure}              
    \begin{subfigure}[t]{0.50\textwidth}
        \centering
        \includegraphics[width=0.6\textwidth]{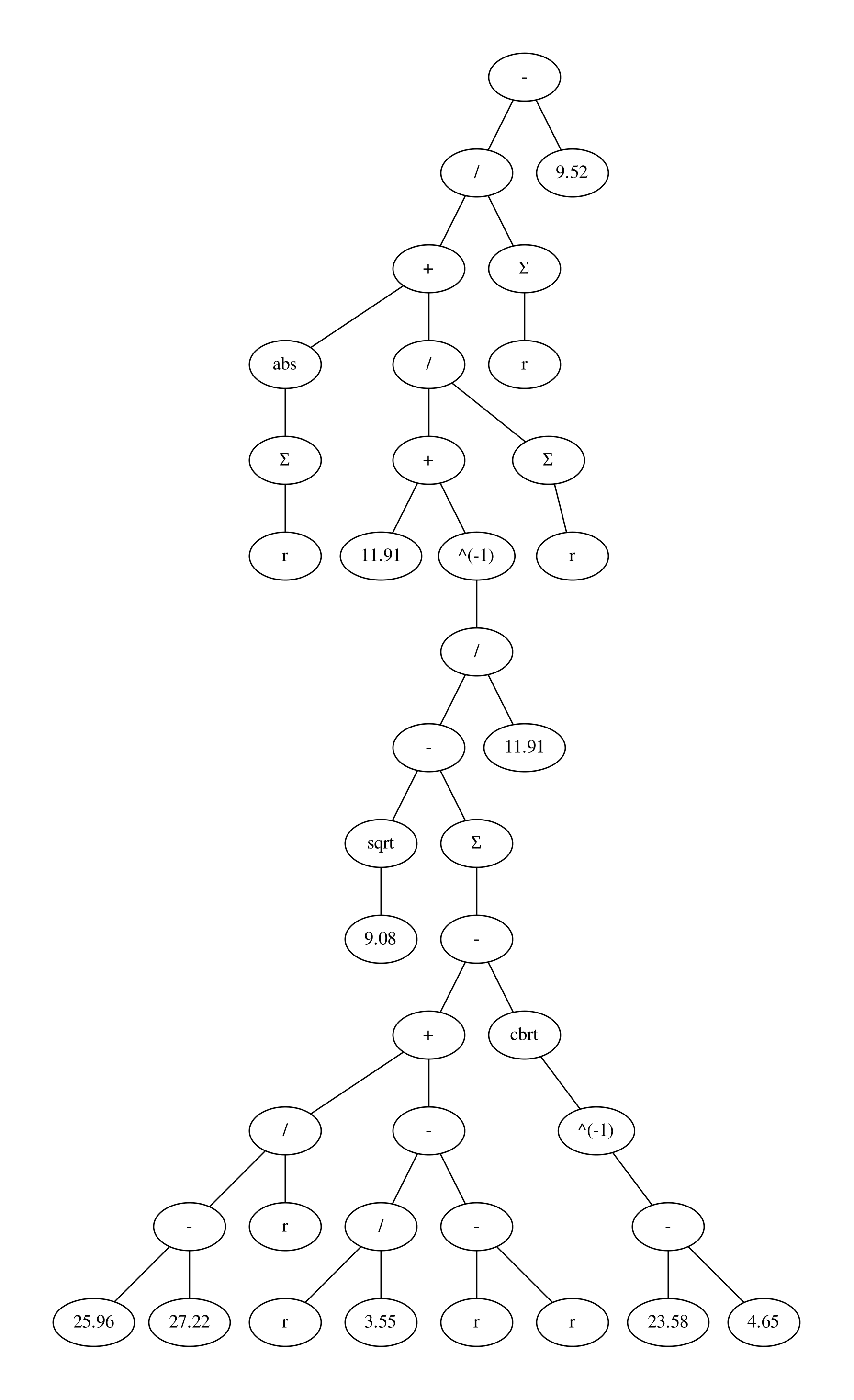}
        \caption{GP5}
        \vspace{-3pt}
        \label{fig:GaB2N3_predict2}
    \end{subfigure}     

   \caption{Graphical tree representations of regression models produced by the genetic algorithm. Trees were constrained by GP hyperparameters to a minimum depth of 1 node and maximum depth of 13 nodes, and they had no constraints on minimum or maximum total nodes. However, the total number of nodes (complexity) was assigned a negative weight in the fitness 3-tuple. }
  \label{fig:trees}
\end{figure}

\begin{figure}
  \centering
  \includegraphics[width=0.75\textwidth]{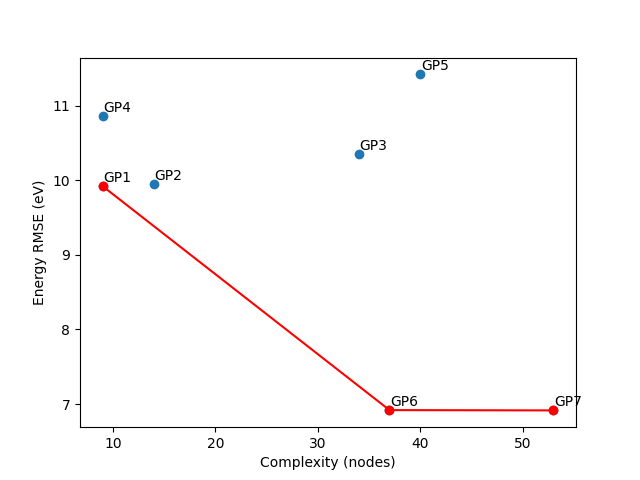}
  \caption{Energy RMSE vs. complexity of regression models generated by the genetic algorithm. Red points along the Pareto front, representing non-dominated individuals, include GP1, GP6, and GP7.}
  \label{fig:pareto_front}
\end{figure}

The genetic algorithm produced several viable two-body carbon interatomic potentials with low computational complexity, shown in Table~\ref{tab:regmodels}. The Pareto front consisting of non-dominated individuals among the potentials developed here includes GP1, GP6, and GP7, shown in Figure~\ref{fig:pareto_front}. Each of these potentials developed similar kernels involving the inversion of a summation over neighboring atomic distances with a constant scaling parameter. The tree graphs of individuals along the Pareto front are shown in Figure~\ref{fig:trees}.

GP1, GP2, and GP4 follow a similar form of a constant $C_1$ divided by a summation over the interatomic distances $r_{ij}$ subtracted from some constant factor $C_2$ minus another constant factor $C_3$. This approximate form is shown in equation~\eqref{eq:potential_form_1}.

\begin{equation}
    \label{eq:potential_form_1}
    E_i = \frac{C_1}{\sum_j (C_2 - r_{ij})} - C_3
\end{equation}

The major notable difference is that GP2 additionally has a summation over pairs of neighbors in the numerator.

GP3 and GP5 follow a form roughly similar to the one provided above which also contains the inversion of a summation over interatomic distances minus a constant factor, but with added complexity as the maximum allowed depth of trees was increased for these models. In these cases, essentially null terms like the one shown in equation~\eqref{eq:gp3_term}, drawn from GP3, arise as a byproduct of the generational process.

\begin{equation}
    \label{eq:gp3_term}
    \left( \sum_j \frac{1}{r_{ij}} \right)^2 \approx 0
\end{equation}

Such terms are a necessary consequence of the evolutionary algorithm and can be safely removed from the model without impacting overall performance. Thus high-complexity GP-based models can often be simplified to achieve a lower complexity with equal accuracy.

GP6 and GP7 were also trained with a large maximum tree depth and were additionally fitted to a more homogeneous training set. The resulting kernels bare little resemblance to those listed above or each other, likely as a consequence of their over-fitting to the training set. This is a testament to the necessity of diverse training data for the GA to perform optimally. Regardless, GP6 and GP7 generalized somewhat well in their performance on the testing set.

\subsection{Model optimizations}

All of the generated potentials along the Pareto front expand its complexity axis, but still have room for improvement in their accuracy. We suspect that the primary limiting factors of these potentials' accuracy are:

\textbf{(a)} Incomplete primitive set
    
Without terms for three- and many- body interactions, or consideration of non-local interactions, the potentials will be inherently limited in accuracy. The primitive set specified here contains a terminal value for pairs of neighboring atoms ($r_{i}$), but lacks terminals for three-body (triplets) or many-body terms, making the consideration of these interactions impossible given the current schema. Now that the ability of GP-based symbolic regression to generate interatomic potentials with high generalizability has been established, the next step is to incorporate triplets and many-body terminals into the primitive set.

\textbf{(b)} Suboptimal hyperparameters

As mentioned in section~\ref{sec:gpoverview}, our hyperparameters were not systematically optimized. Rather, they were derived heuristically based on current genetic programming literature. Optimal hyperparameters are important to ensure maximum performance from the evolutionary process, and the use of systematically optimized hyperparameters would likely improve the genetic algorithm's speed and the performance of generated individuals. Optimizing hyperparameters for genetic algorithms is uniquely challenging due to the resources and time required to run even a single GA instance to completion using any significant number of generations and population size. Additionally, evaluating the performance of a given instantiation of hyperparameters is non-trivial due to the inherent randomness in GAs, which may cause a sub-optimal set of hyperparameters to produce better individuals than a more optimal hyperparameter set on some random validation set.

Addressing these two limitations in the current model is crucial to obtaining the near-DFT accuracy exhibited by HDNNPs and Gaussian approximation potentials.

\begin{table}[ht]
    \centering
    \caption{Regression models produced by the genetic algorithm. Each model's fitness is characterized by its RMSE (eV/atom) relative to target DFT calculations and complexity (number of nodes). Non-dominated individuals include GP1, GP4, GP6, and GP7.}
    \def\arraystretch{5}
    \begin{tabular}{|p{1.5cm}|p{1.75cm}|p{1.75cm}|p{10.5cm}|}
     \hline
     \textbf{Name} & \textbf{RMSE (eV)} & \textbf{Complexity (nodes)} & \textbf{Regression model}\\
     \hline
     GP1 & 9.923 & 9 & $E_i = \frac{1}{\sum_j \left(2.312 - r_{ij}\right)} - 8.48$\\
     GP2 & 9.951 & 14 & $E_i = \frac{\sum_j r_{ij}}{\sum_j \left(9.85 - r_{ij}\right)} - 9.071$\\
     GP3 & 10.352 & 34 & $E_i = \sqrt[3]{\frac{1}{\left(\sqrt[3]{\sum_j \frac{r_{ij}}{19.83}} - 1.663\right)\left(\left(\sum_j \frac{1}{r_{ij}}\right)^2 - 1.679\right)} - 25.47}$\\
     GP4 & 10.861 & 9 & $E_i = \frac{5.54}{\sum_j (4.65 - \sum_j r_{ij})} - 8.5$\\
     GP5 & 11.421 & 40 & $E_i = \frac{\sum_j r_{ij} + \frac{11.91 + \frac{11.91}{3.013 - \sum_j \left(\frac{r_{ij}}{3.55} - \frac{1.26}{r_{ij}} - 0.375\right)}}{\sum_j r_{ij}}}{\sum_j r_{ij}} - 9.52$\\
     GP6 & 6.919 & 37 & $E_i = \frac{\sum_j r_{ij}}{478.297} + \sqrt[3]{\frac{8.66}{9.63 - (\sum_j 1)^2}} + \frac{1}{4.66 - \sum_j r_{ij}} - 8.636$\\
     GP7 & 6.915 & 53 & $E_i = 5.682 \left[\left(51.88 + \abs{\frac{1}{\sum_j (\frac{\sqrt[3]{\frac{29.91}{r_{ij}}}}{489.429} - \frac{r_{ij}^2}{738.499})}}\right) \sqrt[12]{\abs{\frac{1}{\left(\sum_j r_{ij}\right)^2} - 0.174}}\right]$\\
     \hline
    \end{tabular}
    \label{tab:regmodels}
\end{table}

\FloatBarrier
\section{Discussion and future work}
\label{sec:future_work}

The GP7 regression model displays the highest accuracy of the potentials produced here, calculating energy within $\pm 7.70$ eV of the DFT target values.

The regression models produced here currently consider only two-body interactions between neighboring atoms to calculate local energy contributions, representing a minimum viable two-body potential. As previously discussed, typical GAPs contain terms for two-, three-, and many-body interactions. This difference in the range of interactions considered by each model could be a major source of the current disparity in energy calculation accuracy.

Future works may explore the expansion of the GP primitive set to contain a terminal for triplets to represent the three-body interaction term, as well as terminals for the many-body interaction term. The analogous kernel-based regression models demonstrated substantial improvement with the introduction of three- and many-body terms, so their introduction in the context of GP-based potentials is likely to have a similar effect. The complexity of generated potentials will necessarily increase as a consequence of considering three- and many-body interactions, but the additional computational complexity is a necessary cost to attain near-DFT accuracy.

Future works may also include programmatic hyperparameter optimization, the other major limiting factor in the current framework. We are currently using holdout validation with four unique partitions for data validation. One method to achieve programmatic hyperparameter optimization is to switch to $k$-fold cross-validation, which would substantially increase time to train the model.

\section{Conclusion}
\label{sec:conclusion}

The genetic programming framework established here produces accurate regression potentials with high correlation to DFT energy calculations. Due to the highly transferable nature of genetic programming algorithms, our framework is equally effective on energy datasets for all atomic structures and can scale directly with the size of any DFT dataset.

The framework can be expanded to include local energy contributions from three- and many-body interactions by adding new terminals for triplets of atoms and individual atomic coordinates. Based on related research on kernel-based regression potentials, we find it highly likely that such additions to the primitive set would result in substantially increased accuracy at the acceptable cost of additional computational complexity. Once three- and many-body interactions are successfully accounted for and a more robust GP-based potential is developed, it will likely be nearer to GAP1 and GAP2 on the Pareto front (with respect to both accuracy and complexity) than other GP potentials produced here.

Our GP-based regression potentials provide a low-complexity alternative to computationally expensive ab initio methods, featuring substantially reduced complexity compared to GAP regression models. However, the potentials produced here strictly consider two-body interactions and do not account for three-body, many-body, or non-local interactions, limiting the maximum accuracy currently obtainable by the models.

We developed several accurate and generalizable carbon potentials through GP-based symbolic regression using a diverse dataset of carbon structures and multiple train-test partitions. We have demonstrated the viability of GP-based potentials as computationally inexpensive interatomic potentials with high generalizability, but the upper bounds of their accuracy remain to be seen.

\section{Code availability}
\label{sec:code_availability}
The code referenced in this paper is open source and available at \href{https://github.com/usccolumbia/mlpotential}{https://github.com/usccolumbia/mlpotential}.

\section{Data availability}
\label{sec:data_availability}
The carbon DFT dataset referenced in this paper is open source and available at \\ \href{https://github.com/usccolumbia/mlpotential/tree/main/data}{https://github.com/usccolumbia/mlpotential/tree/main/data}

\bibliographystyle{unsrt}  
\bibliography{references}

\section{Contributions}
This research was conceived and managed by M.H. and J.H. The carbon DFT dataset were provided by A.R. The code was written by A.E with the guidance of J.H. The manuscript was drafted by A.E. and revised by J.H. and M.H.

\section{Conflicts of interest}
The authors declare no conflicts of interest.

\section{Acknowledgements}
\label{sec:acknowledgements}
Research reported in this work was supported in part by NSF under grants 1940099, 1905775, and 2110033. The views, perspective, and content do not necessarily represent the official views of NSF. This research was supported in part by a grant from the Magellan Scholarship program through the University of South Carolina's Office of Undergraduate Research.

\end{document}